\documentclass[10pt,preprint]{aastex}
\usepackage{natbib}
\usepackage{amsmath,amsfonts,amssymb}
\usepackage{graphicx}
\def\deg{{\rm o}}
\def\idm#1{{\mbox{\scriptsize #1}}}

\newcommand\Ym{\langle Y\rangle}
\newcommand\Chi{{(\chi^2_\nu)^{1/2}}}
\def\astrobj#1{#1\ }

\def\url#1{\texttt{#1}}
\newcommand\pstar{{\astrobj{HD~169830}}}
\newcommand\pstaro{{\astrobj{HD~12661}}}

\begin{document}
%
%
%
\title{The exosystems around HD~169830 and HD~12661: 
   are they dynamical twins?}
   
\author{Krzysztof Go\'zdziewski\altaffilmark{1}}
\affil{Toru\'n Centre for Astronomy, N.~Copernicus University,
Gagarina 11, 87-100 Toru\'n, Poland}
\author{Maciej Konacki\altaffilmark{2}}
\affil{Department of Geological and Planetary Sciences, California
Institute of Technology, MS 150-21, Pasadena, CA 91125, USA \\
Nicolaus Copernicus Astronomical Center, Polish Academy of Sciences,
Rabia\'nska 8, 87-100 Toru\'n, Poland}

\altaffiltext{1}{e-mail: k.gozdziewski@astri.uni.torun.pl}
\altaffiltext{1}{e-mail: maciej@gps.caltech.edu}
  
\begin{abstract}
The new 2-planetary system around \pstar has been announced  during the XIX-th IAP 
Colloquium "Extrasolar Planets: Today \& Tomorrow" (Paris, June 30 - July 4, 2003)
by the Geneva Extrasolar Planet Search team.  We study the orbital dynamics of 
this system in the framework of the $N$-body  problem. The analysis of its orbital 
stability is performed using the long-term integrations and
the fast indicators,  the Mean Exponential Growth factor of Nearby Orbits and
the Frequency Map Analysis.  The \pstar appears to be located in a wide
stable region of the phase space.  The ratio of the mean motions of the
planets \pstar b and c is between low-order mean motion resonances, 9:1 and
10:1. The long-term integration  of the coplanar configurations, conducted
over 1~Gyr, reveals that the  eccentricities of the companions vary with a
large amplitude $\simeq 0.4-0.5$ but there is no sign of instability.  The
orbital parameters of the planets resemble those of another
2-planetary system, around HD~12661. Both of them can be classified as
hierarchical  planetary systems. We investigate whether these two exosystems
are dynamically similar. Such similarities may be important for finding out 
if the formation and subsequent orbital evolution of exoplanetary systems 
obey common rules. 
\end{abstract}
%
%
\section{Introduction}
%
%

The radial velocity survey for extrasolar planetary systems conducted by  the
Geneva Extrasolar Planet Search Team has recently revealed a new planetary
system  around HD~169830. This discovery has been announced during the XIX-th
IAP Colloquium "Extrasolar Planets: Today \& Tomorrow", (Paris, June 30 -
July 4,  2003). In this paper we perform a preliminary analysis of the
orbital stability  of the \pstar system and its dependence on the orbital
parameters that are unconstrained by radial velocity (RV) observations.  In
the numerical experiments, we use two sets of the orbital parameters from 
the 2-Keplerian fits to the RV measurements. The first
set was published by the authors on their web site in June 2003
\footnote{\url{http://obswww.unige.ch/~udry/planet/hd169830\_syst.html}}
and the second, most recent set is from \cite{Mayor2003}. Our early analysis of the 
first set of the orbital elements revealed a curious, close similarity to
the orbital elements of the HD~12661 system \cite[]{Fischer2003,Gozdziewski2003d}. 
These two hierarchical planetary systems (HPS) appeared to have very similar ratios 
of the minimal masses $\mu=m_{\idm{b}}/m_{\idm{c}} \simeq 1.3$, the  
semi-major axes $\alpha=a_{\idm{b}}/a_{\idm{c}}\simeq 0.3$ and similar initial 
eccentricity of the outer planet $e_{\idm{c}} \simeq 0.3$ with simultaneously small
eccentricity of the inner planet $e_{\idm{b}} \simeq 0$. As we show
in this paper, in such a case both systems would be located in a large 
island of the secular apsidal resonance with the semi-major axes antialigned in the
exact resonance. There could be another planetary system, around HD~160691 
\cite[]{Jones2002b}, with a qualitatively similar dynamics as suggested by 
\cite{Gozdziewski2003e}. In a relatively small number of about 14 multi-planetary 
systems known to the date\footnote{\url{http://www.encyclopaedia.fr}}, these three 
cases of supposedly identical configuration inspired us to ask whether such a dynamical 
resemblance can be casual or has a deeper origin in their formation scenario and/or 
orbital evolution.

However recently, the Geneva team has published a new orbital solution substantially 
different than the first one \cite[see][]{Mayor2003}. In the new solution the 
elements of the outer companion are
dramatically altered. The period and the semi-major axis of the outer planet are 
changed by $\simeq 600$~d and $\simeq 0.7~$AU respectively. This obviously leads 
to a qualitatively different orbital configuration than the one analyzed before.
The proximity of the \pstar and \pstaro systems does not seem to be so
clear anymore. Still, the orbital elements of the inner companions
appear to be almost identical in the two planetary systems. Because
the observational window of \pstar covers only about 1~orbital period of the outer
planet, its orbital elements are still not well constrained. In 
our numerical experiments, we use both solutions although the old one
essentially for a reference and an interesting comparison with the predictions given
by the secular octupole-level theory~\citep{Lee2003}. 

According to \cite{Lee2003}, one has to employ a proper representation of the 
orbital fits to the RV observations, particularly 
in the case of the HPS. The orbital fits to the RV of HPS  are best interpreted in
Jacobi coordinates and these coordinates should be used in the dynamical 
investigations. Specifically, the authors demonstrate that the orbital
time-evolution when expressed in the commonly used astrocentric elements can
lead to spurious effects like a large scatter of the osculating semi-major
axis and eccentricity of the outer planet about the secular value or a
significant dependence of the integration  results on the initial epoch. 
Unfortunately, the RV measurements of the \pstar system have not been
published to the date. However, assuming that the best-fit parameters  
$K,n,e,\omega,T_{\idm{p}}$ (where, for every planet involved, $K$ is the 
semi-amplitude of RV variations, $n$ is the mean motion, $e$ is the eccentricity, 
$\omega$ is the periastron argument and $T_{\idm{p}}$ is the time of periastron passage) 
describe a 2-Keplerian fit, they represent in fact the orbital parameters in the 
Jacobi coordinate system \citep{Lee2003} and using them, we can recover 
the semi-major axes $a$ and the minimal masses of the companions. 
In some of the numerical experiments carried out in this paper, the inclination of 
the outer planet was changed by a tiny value, to 89.9$^{\circ}$, to allow the 
system enter the third spatial dimension for the purpose of the numerical integrations. 
The periastron passage of the inner planet is selected as the initial epoch of the 
osculating elements inferred from the 2-Keplerian solutions. The two ICs used in the
computations are given in Table~\ref{tab:tab1}. The first, preliminary one, IC is labeled 
by ICI and the current one by ICII. In some experiments we modified ICI,
by setting $e_{\idm{c}} \simeq 10^{-6}$, to avoid numerical singularities.
The elements of the \pstaro system
are given in Table~\ref{tab:tab3} and we refer to them as IC0 throughout this
work.

\section{Numerical setup}

We use the so-called fast indicators, MEGNO and FMA as the main numerical tools to 
study the orbital stability of the \pstar system. The indicator
called MEGNO (the Mean Exponential Growth Factor of Nearby Orbits), has been
invented by \cite{Cincotta2000} and we have applied it  in the studies of the
exosystems' dynamics described in a series of recent papers \citep[for
details and references on MEGNO
see, for example,][]{Gozdziewski2001a,Gozdziewski2003d}.     The MEGNO is a very efficient
tool---typically,  it makes it possible to distinguish between regular and
chaotic dynamics during  an integration of the system carried out over
$\simeq 10^4$ orbital periods of the outermost planet.  This indicator provides
a direct estimate of the stability in the strict sense of the  maximal
Lyapunov Characteristic Number (LCN). As in the previous works, the MEGNO
integrations are driven by a Bulirsh-Stoer integrator, namely the ODEX
code \citep{Hairer1995}. The relative and absolute accuracies of the
integrator are set to $10^{-14}$ and $5\cdot 10^{-16}$, respectively.
The zones of stability, corresponding to  quasiperiodic motions of the
system, are marked in the MEGNO maps by values close to 2.  The positions of
the nominal conditions are marked in the contour plots by the intersection 
of the two thin lines. In this work, we make yet another use of the integrations.
During the MEGNO integrations we  simultaneously
analyze the osculating elements to study the short-term dynamics of  the
system.  The
osculating  orbital elements feed the Frequency Map Analysis (FMA) algorithm
\citep{Laskar1993}.  The FMA, an already classic fast indicator, allows us to
detect orbital resonances  and determine diffusion rates of the
fundamental frequencies in the planetary system
\citep{Robutel2001}. The diffusion rate is a quantity that cannot be
determined by the LCN, however it also measures the lack of regularity of an
orbit.  Along a quasi-periodic solution the fundamental frequencies are
constant.  For a chaotic orbit these frequencies change and by calculating
their diffusion rates, one can directly detect the macroscopic changes of the
orbital elements. Thus by combining both methods, it is possible to derive
extensive information on  the system's dynamics.  
 
Since our computations concern mostly a short-term dynamics of the
planetary system, the FMA is very useful in identifying the positions
of the mean motion resonances (MMRs). In order to identify these resonances,
during every integration of MEGNO, the complex functions
$
     f_k(t_i) =  a_k \exp \mbox{i}\lambda_k(t_i)
$
are computed for planets $k=\mbox{b,c}$ at the discrete times $t_i$,
where $t_{i+1}-t_{i} < 0.5 P_{\idm{b}}$, over the time $2T$ of the order of
$10^4~P_{\idm{c}}$ ($P_{\idm{b,c}}$ are the orbital periods of the inner and
outer companion, respectively). Here, $a_k$ denotes the semi-major axis and
$\lambda_k$ is the mean longitude of the planet. If the  motion is
quasiperiodic then in the time series $\{f_k(t_i)\}$, the frequency $\nu_k$ 
corresponding to the largest amplitude $a_k^0$ {\bf }  is one of the
fundamental orbital frequencies of the system, called the proper mean motion,
$n_k$ \citep{Robutel2001}.  Because the secular  frequencies are much smaller
than the orbital frequencies, in the first  approximation the precession of
the orbits can be neglected and MMRs can be identified through the condition
$
       q \nu_{\idm{b}} - p \nu_{\idm{c}} \simeq 0,
$
where $p,q>0$ are prime integers. The FMA gives the ratios
$\nu_{\idm{b}}/\nu_{\idm{c}} \simeq p/q$,  
and then $p$ and $q$ can be easily resolved by the continued fraction
algorithm.

After finding $\nu_{\idm{b,c}}^{(1)}$ and $\nu_{\idm{b,c}}^{(2)}$ where
$\nu_{\idm{b,c}}^{(1,2)}$ are the proper mean motions $n_{\idm{b,c}}$ 
obtained after 
integrating the system over the time intervals $[0,T]$,
and $[T,2T]$, respectively,
 the diffusion rate is determined through   
$
   \sigma_k = 1 - \nu_k^{(2)}/\nu_k^{(1)}
$
\citep{Robutel2001}. 
Following these authors, if the motion is quasiperiodic,
$\sigma_k$ is equal
to zero or,  computed with the FMA algorithm, is a very small quantity, 
typically less than 
$\simeq 10^{-8}$ while for an irregular chaotic motion, $\sigma_k$ is larger by several
orders of magnitude. The inferred diffusion of the semi-major axis , can be
derived through the Kepler law $n^2 a^3 = \mu$ as $\Delta n/n \simeq -3/2
(\Delta a/a)$.  In this way, although the time span of the integrations 
is relatively very short, the most relevant instabilities of the motion can
be detected and identified.
   
In this paper, we use the FMA code kindly provided by David Nesvorny on his
web page\footnote{\url{http://www.boulder.swri.edu/~davidn/fmft/fmft.html}}.
According to the author, the code incorporates the idea of the Frequency 
Modified Fourier Transform, FMFT \citep{Nesvorny1997}. We use a variant of
the FMFT, with the so-called additional non-linear correction (see the cited 
paper for details).  We have tested this code to make sure that it resolves
the mean motions of non-interacting (Keplerian) orbits with the accuracy
consistent with the internal precision of the method (set to $10^{-10}$).

\section{Orbital fits}

The two distinctively different orbital solutions, ICI and ICII, announced within 
a short period of time by the Geneva group strongly suggest that at least one of 
them corresponds to a local minimum of $\Chi$. Unfortunately, it is not
possible to verify their elements since the RV measurements are
not available. In order to get at least some insight into the reliability
of the newest set of orbital parameters, we have digitized the figure with
RV measurements published on their WWW site 
\footnote{\url{http://obswww.unige.ch/~udry/planet/hd169830\_syst.html}}.
The 93 data points we obtained differ from the real observations (for example, 
it is difficult to recover the exact moments of the observations),
 but they still 
properly describes the overall shape of the observed RV curve and its characteristic 
features. Note also that our set of "measurements"\footnote{The set is 
available upon request from KG} is smaller than the real one (112) 
because in some parts of the RV curve it was very difficult
to clearly resolve all points due to a dense sampling of the real measurements.
We assume that the observational errors are given by $\sigma =
\sigma_{\idm{0}} + \sigma_{\idm{1}}$ where $\sigma_{\idm{0}}=8.5$~m/s and 
$\sigma_{\idm{1}}$ is an artificial normal random noise of 0~mean and 
the standard deviation about $1$, resulting in the "errors" in the range
$[6,11]$~m/s and roughly approximating the scale of the real errors. 

We used the genetic algorithm scheme (GA) implemented by \cite{Charbonneau1995} 
in his publicly available
\footnote{\url{http://www.hao.ucar.edu/public/research/si/pikaia/pikaia.html}}
code PIKAIA to to look for a global, coplanar orbital solution. The data were modeled 
with 2-Keplerian model~\citep[as in][]{Gozdziewski2003e} and the self-consistent
Newtonian model, i.e., driven by the full $N$-body dynamics~\citep{Laughlin2001}.
For the first model the GA code was executed a few hundreds of times and
repeatedly found a solution very close to the best one, given in Table~\ref{tab:tab2}.
Even though the number of our "measurements" is smaller than in the real set,
the similarity of the orbital solutions is very significant. Remarkably, the GA found 
very similar parameters for the outer planet (although the differences are larger
than for the inner planet). The best fit parameters have the $\Chi \simeq 1.21$ and 
the rms of $\simeq 10$~m/s. This experiment was also repeated for
two other data sets with 104 and 105 points (from two slightly different digitization) 
with the errors in the range $[5,12]$~m/s and $[7,10]$~m/s, respectively. 
The obtained 
best fit solutions were qualitatively the same as those given in Table~\ref{tab:tab2}
(let us note that the differences were most significant for 
$e_{\idm{b}}\simeq 0.29$,
$e_{\idm{c}}\simeq 0.29$
and $\omega_{\idm{c}} \simeq 270^{\circ}$).
Our analysis indicates that the ICII by the Geneva group
is really close to the global minimum of $\Chi$ and should not change at least if based
on the currently available set of the RVs.

With the second model that accounts for the mutual interactions between planets,
the best fit solution is not significantly better. It is characterized by the $\Chi \simeq
1.21$ and the rms of $\simeq 9.5$~m/s. The corresponding orbital parameters are
very similar to those from the 2-Keplerian fit (see Table~\ref{tab:tab2}; for a
reference, the  synthetic RV-curves and the digitized "data" points are also shown in
Fig.~\ref{fig:fig4}). This result suggests that the current RV set does not
allow to detect significant $N$-body coupling (for example, from low-order 
resonances). Finally, to roughly estimate the formal errors of the self-consistent 
Newtonian fit, we determined the confidence intervals of $\Chi$~\citep{Press1992}. 
The results are illustrated in Fig.~\ref{fig:fig5}. Clearly, the formal errors of 
$a_{\idm{c}}$ and $e_{\idm{c}}$ (semi-major axis and eccentricity of the outer planet) 
are substantial but the minimum of $\Chi$ is well localized. We note that to estimate 
the conservative errors in the best-fit parameters, the effect of stellar "jitter" 
should be taken into account \citep[as in][]{Gozdziewski2003d}.

\section{Orbital stability of the \pstar system}

Both orbital configurations of \pstar, ICI and ICII, appear to be stable. The
evolution of the Jacobi osculating orbital elements computed over the time
span  of $\simeq 3.6 \cdot 10^{4}$ revolutions of the outer planet (about
$150,000$~yr) for the ICI fit is illustrated in Fig.~\ref{fig:fig1}. During
this time, the eccentricities  (see Fig.~\ref{fig:fig1}c)  vary with a large
full-amplitude of about $0.35$. Fig.~\ref{fig:fig1}d reveals a presence of
the  secular apsidal resonance (SAR) in which  the apsides are on the average
antialigned. The semi-amplitude of the librations, $\theta$ (where 
$\theta=\varpi_{\idm{b}}-\varpi_{\idm{c}}$ and $\varpi_{\idm{b,c}}$ are the
respective longitudes of periastron), is about $90^{\circ}$.  Apparently, the
orbital elements vary in a regular way and this is strictly confirmed by the
MEGNO signatures.  The temporal changes of MEGNO, $Y(t)$, are shown in
Fig.~\ref{fig:fig1}e and its mean value, $\Ym(t)$ (Fig.~\ref{fig:fig1}f),
perfectly converges to~2. For a comparison, we computed these quantities for
the \pstaro system using the IC0 from Table~\ref{tab:tab3}. The results are
shown in Figure~{\ref{fig:fig3}}. In both these cases, the variations of the
orbital elements and the character of MEGNO convergence are qualitatively
identical.   However, the ICII fit for the \pstar system results in a
qualitatively different orbital evolution (Fig.~\ref{fig:fig2}). The SAR does
not seems to be present anymore. Nevertheless, the MEGNO signature (computed
over about $550,000$~yr) indicates a stable, quasiperiodic configuration.

The MEGNO signatures have been verified by  direct  1~Gyr integrations using
the RMVS3 integrator from the SWIFT package \cite[]{Duncan1994}. We repeated
the integrations with two different time steps, equal to 8 and 10~days. 
As one would expect, no instability occurs during this
time and the orbital  elements vary within the bounds determined by the
short-term integrations. We do not show these results as they are basically
an extension of the plots shown in Figures~\ref{fig:fig1}, \ref{fig:fig2} and
\ref{fig:fig3}.  Obviously, due to the uncertainties of the orbital fits,
such an examination  of the isolated IC's is not representative for the
system's dynamics.  In order to find out whether the dynamics  is robust to
small adjustments of the initial condition, we have computed one-dimensional
scans of $\Ym$ by changing the semi-major axis of the outer planet and
keeping the other orbital parameters fixed at their initial values given in
Table~\ref{tab:tab1} for \pstar and \ref{tab:tab3} for \pstaro. The results of this experiment are
shown in Fig.~\ref{fig:fig6}. The MEGNO scan (the upper graphs of the
respective panels in Fig.~\ref{fig:fig6}), computed with the resolution of 
about $3\cdot 10^{-4}$~AU, reveals a number of spikes.  Most of these spikes
represent the MMRs between the planets. This identification is based on the
FMA as described in the previous Section.  We marked MMRs of the order  $p+q$
not grater than about 20. These scans reveal that the dynamical environment
of the nominal ICs is qualitatively the same for the ICI of the \pstar and the IC0 of
\pstaro systems.  The outer planet lies in a stable zone between strong MMRs 6:1 and
7:1. The graph for the ICII of the \pstar system reveals a different dynamical environment.
Clearly,  the change in $e_{\idm{c}}$ (compared to $e_{\idm{c}}$ from ICI) "shifted" the 
system to the zone between the 9:1 and 10:1 MMRs.

The FMA working with the MEGNO code enabled us to find
out how sensitive both methods are to the instabilities of the motion. Bottom
parts of the panels of Fig.~\ref{fig:fig6} show the diffusion rate
computed  for the outer planet and estimated over its $\simeq 18,000$ orbital
periods.   In the regions of quasiperiodic motions, as classified by MEGNO,
the diffusion  rate, $\sigma_{\idm{c}}$,  is smaller than about $10^{-8}$.
Such small values  ensure  us that the motion is close to a quasiperiodic
evolution. In the  MMRs zones $\sigma_{\idm{c}}$ grows
up to $10^{-2}-10^{-1}$. Hence in these areas, the  osculating orbital
elements exhibit macroscopic changes. Both algorithms are in an excellent
accord, as they provide the same positions of the MMRS and very similar 
estimates of their widths. In other experiments, we noticed that for much
shorter  integration times, $\simeq 0.45\cdot10^4 P_{\idm{c}}$, both
algorithms can still  detect all the relevant MMRs although in this case
they  do not allow to point out clearly enough some of the weak resonances. 
In fact, for  the four times longer integration ($\simeq 1.8\cdot10^4 P_{\idm{c}}$), 
the FMA seems to be even
more sensitive to the  presence of weak resonances than the MEGNO algorithm
is. 
We note here that by a compromise
forced by the numerical  efficiency of the code,  in further runs of MEGNO,
we  have set the integration time to  about $0.9\cdot10^4 P_{\idm{c}}$.

\section{Global dynamics of the \pstar system}

To extend the above one-dimensional analysis, we can investigate 
the stability of the \pstar system in a few
representative  planes of its orbital parameters.
 
The program computing MEGNO simultaneously evaluated the maximal values of
the eccentricities, $e_{\idm{b,c}}^{\idm{max}}$,  attained during the
integration time. We also stored the maximal value of semi-amplitude of the
librations, $\theta^{\idm{max}}$,  after every step (set to $P_{\idm{c}}$) of 
renormalization of the variational equations. It
helped us to detect the apsidal resonance and to estimate the semi-amplitude
of the librations. The maximal value of the critical argument $\theta$ was
taken relative to the center of the libration  $0^{\deg}$ or $180^{\circ}$.
To avoid the effects of a possible transition into the SAR,  the
determination of $\theta^{\idm{max}}$ was started after the first half of the
integration period (about $0.45 \cdot 10^4 P_{\idm{c}}$).  Finally, if
$\theta^{\idm{max}}<90^{\circ}$, then we treated this value as a
semi-amplitude of the apsidal librations.  The period of the integrations is
relatively short but as we show in 
\cite{Gozdziewski2003c}, such information on the short-term dynamics can still
give us much insight into the global behaviour of the system.
The results are illustrated in Figs.~\ref{fig:fig7}--\ref{fig:fig9}, 
where the left panels are for MEGNO, $\Ym$, the middle panels are for
$\theta^{\idm{max}}$ and the right panels are for 
$e_{\idm{b}}^{\idm{max}}$. 
In these scans, the initial parameters that are varied are the coordinates of the maps
and the other initial orbital elements  are fixed at
their nominal values given in Table~\ref{tab:tab1}.

Figure~\ref{fig:fig7} is for the $(a_{\idm{c}},e_{\idm{c}})$-plane. 
Since the resolution of these maps is $200 \times 50$ data points and the
integration time is shorter than that of the one-dimensional $a_{\idm{c}}$
scan (shown in Fig.~\ref{fig:fig6}), there is a lack of some fine resonance structures visible in
Fig.~\ref{fig:fig6}. Yet the location of the  dominant low-order MMRs: 6:1,
13:2
and 7:1 for ICI and 8:1, 9:1 and 10:1 for ICII,  as well as their widths
are clearly marked. The system would be chaotic for $e_{\idm{c}}$   roughly
grater than 0.5-0.6. In the central, stable parts of the MMRs, the maximal
values of $e_{\idm{b}}$ 
are small indicating their stabilizing influence on the
motion.  For ICI of the system around \pstar (as well as for IC0 of the \pstaro system,
not shown here), the plot for
$\theta^{\idm{max}}$ reveals an extended zone of the SAR about the libration
center of $180^{\circ}$.  For very small initial $e_{\idm{c}} \simeq 0$, the
semi-amplitude of librations  reaches the limiting $90^{\circ}$. If the
eccentricity becomes larger than $\simeq$ 0.005 then the semi-amplitude
decreases rapidly to $60^{\circ}$-$70^{\circ}$. The same
type of $(a_{\idm{c}},e_{\idm{c}})$-scans of MEGNO for the HD~12661 system is
shown in \cite{Gozdziewski2003c} (his Fig.~6) and \cite{Gozdziewski2003d}
(their Fig.~3).  In both cases, the MEGNO structures around the nominal ICs
are very similar. Interestingly, the $(a_{\idm{c}},e_{\idm{c}})$-maps of
MEGNO for some of the fits to the RV data of the HD~160691 planetary system
in \cite{Gozdziewski2003e} (e.g., the fits GM5 and GM6, Fig.~5 in that
paper), describe a dynamical setup very similar to these of the \pstar (with ICI)
and \pstaro systems. This is no longer true for the ICII of the \pstar system. 
In this case, the
zone of the SAR is confined to a very small $e_{\idm{c}}$ (but over wide range
of $a_{\idm{c}}$) and to the centers of the MMRs (see Fig.~\ref{fig:fig7},
bottom panels). 

Figure~\ref{fig:fig8} shows the behaviour of MEGNO, the semi-amplitude
of  apsidal libration and the maximal eccentricity of the inner planet for
\pstar system on the $(e_{\idm{b}},e_{\idm{c}})$-plane. The nominal
configuration of the  system (denoted as previously  by the intersection of
the two thin lines) is located in an extended  zone of quasiperiodic motions
as seen in the MEGNO map (Fig.~\ref{fig:fig8},  panels 
in the left column). The $e_{\idm{b}}^{\idm{max}}$-maps 
(Fig.~\ref{fig:fig8}, right column)  enable
us to determine the border of the global instability established by these
values of the initial $e_{\idm{c}}$ which lead to large maximal
eccentricities. For ICI, this border is substantially  shifted towards larger
initial eccentricities compared to the border of chaotic zone visible in the
MEGNO map but overall it has a similar shape. The same effects are seen for
ICII. In the SAR-maps (Fig.~\ref{fig:fig8},  middle column), for both IC's  one can
find a very sharp border of this resonance in the region of  small
$e_{\idm{b}}$.  For ICI, it ends at the border of unstable motion present in
the relevant  $e_{\idm{b}}^{\idm{max}}$ panel. For ICII, 
the semi-amplitude of the librations  is about $60^{\circ}$--$70^{\circ}$ 
in this area but 
as for ICI (Fig.~\ref{fig:fig7},  middle panel), a very narrow  strip
of semi-amplitudes of $\sim 90^{\circ}$ for the initial  $e_{\idm{c}} \simeq 0$
is present. These large-amplitude librations for small $e_{\idm{c}}$ are
explained in the next section. Let us note that also in this test, the
obtained  picture of the stability zones for ICI of the \pstar system
is qualitatively the same as for the
\pstaro system --- see Fig.~5 and the discussion from
\cite{Gozdziewski2003c}. However, the SAR-map is quite different for 
ICII since the SAR is confined to a narrow  zone
of regular motion about small initial $e_{\idm{c}}$ and also to  unstable
zone which is seen around $(e_{\idm{b}},e_{\idm{c}})=(0.47,0.47)$.

In the next set of experiments, we determine the zones of stability while the system
inclination and hence planetary masses are varied. Assuming that the
(unknown) inclinations of the orbits are the same
$i=i_{\idm{b}}=i_{\idm{c}}$, the relative inclination $i_{\idm{rel}}$ of the
orbits depends on the difference of the nodal longitudes, $\Delta \Omega
=\Omega_{\idm{c}}-\Omega_{\idm{b}}$, according to the formula:
\[
\cos (i_{\mathrm{rel}}) = \cos(i_{\idm{b}}) \cos(i_{\idm{c}}) +
\sin (i_{\idm{b}}) \sin(i_{\idm{c}}) \cos \Delta \Omega.
\]
We assume that both orbits are prograde. By changing $i$, the masses are
altered because the minimum masses, $m_{\idm{b,c}} \sin i$, corresponding to
edge-on orbits, must be preserved.

The results are shown in Fig.~\ref{fig:fig9}.  For both ICI and ICII, even 
for large relative inclinations, the zones of stability extend up to very
low system inclinations,  $i \simeq 15^{\deg}$. This is supported by the
$e_{\idm{b}}^{\idm{max}}$-scan which  is shown in the right panel of this
figure.  In some unstable zones, seen in the MEGNO  map  for the ICI
(Fig.~\ref{fig:fig9}, top-left panel), we can identify the 13:2 MMR. It
quickly  destabilizes the system in the areas about $i\simeq 45^{\circ}$---in
these regions  both eccentricities increases rapidly up to 0.8--0.9.  In the strip
about low inclinations, most likely thanks to the SAR, this effect is absent.

For ICI, due to the initial $e_{\idm{c}}\simeq 0$ and a geometrical 
singularity (for very small eccentricity the argument of periastron becomes undefined), 
our numerically computed $\theta^{\idm{max}}$-map 
(top-middle panel of Fig.~\ref{fig:fig9}) does not allow to clearly detect the zone 
of the SAR. However, in our preliminary study of the ICI we set
$e_{\idm{c}} \simeq 0.007$ and then we found a wide zone of the SAR in the central 
part of the scanned area. The overall picture is
qualitatively the same as in the case of the \pstaro 
system~\citep{Gozdziewski2003d} where for low-$i_{\idm{rel}}$ configurations, 
the SAR persists almost in the entire range of the system inclination $i$. 

For ICII, the unstable zones in the MEGNO map are basically similar to those 
of the ICI of the \pstar and \pstaro systems. Interestingly, all  the maps for
maximal $e_{\idm{b}}$ reveal characteristic quarter-circle areas in which the
eccentricity grows up to 1. In the strips near the central parts and
corresponding to  $e_{\idm{b}}\simeq 1$ the configuration becomes chaotic
(compare the MEGNO- and $e_{\idm{b}}^{\idm{max}}$-maps).
The analysis of the orbits in these areas shows that
together with the excitation of the inner planet's eccentricity, the inclination of
this planet oscillates  with a very large amplitude. It is illustrated in
Figure~\ref{fig:fig10} for the ICII and the initial relative inclination set 
to $80^{\circ}$. Let us observe that in this case $e_{\idm{b}}$ approaches
0.9 while the inclination goes up to $140^{\circ}$. Nevertheless, the system
still remains rigorously stable.  This phenomenon seems to be an analogue of the Kozai
resonance~\citep{Kozai1962} that is known for highly inclined  asteroids and
generally has been observed in triple systems in which the mass of the inner body 
is much smaller than the mass of the central and the outer 
body~\citep{Holman1997}.
In the  multi-planetary exosystems, the masses of the
perturber and the inner body are comparable which makes the 
problem more complex than in the asteroidal approximation. 
Let us note that a possibility of 
this phenomenon has been analyzed in $\upsilon$~Andromedae case \citep{Chiang2001}.
The inclined configurations which we study here are rather
special because it has been assumed that the initial inclinations are the
same for both planets. However, a similar behaviour can be expected when the
absolute inclinations are varied independently.  
It has been observed also
in other hierarchical  exosystems 
\citep[see for example][]{Gozdziewski2003b,Gozdziewski2003c}.
Since this phenomenon provides strong dynamical limits on the
relative inclinations, it certainly deserves a detailed
study which we plan to carry out in a future paper.

\section{Secular apsidal resonance}

The SAR in the \pstaro exosystem was analyzed numerically in \cite{Gozdziewski2003c}. 
We found an extended zone of the SAR in wide ranges of the semi-major axes, 
eccentricities and masses of the companions. Similar behavior of the SAR 
can be observed for the ICI of the \pstar system. It should be noted that those 
results were based on the IC of the \pstaro system derived from the 2-Keplerian model by 
\cite{Fischer2003}. Our attempts to analyze the same RV data set 
resulted in a rather different orbital solution \cite[]{Gozdziewski2003d}. 
Yet, the overall SAR features did not change. These results have 
been confirmed and greatly extended analytically  by \cite{Lee2003} in the 
framework of their octopole-level secular planetary theory. 

Here, we literally quote that basics of their theory, which are relevant for 
our further analysis. This theory describes the secular dynamics in  a
coplanar hierarchical 2-planet system. It is assumed that MMRs are absent. 
The equations of motion originate from the Hamiltonian which is expanded up 
to the third order in parameter $\alpha=a_{\idm{b}}/a_{\idm{c}}$  and
averaged  over both mean longitudes $l_j$, $j=\mbox{b,c}$.  The relevant
parameters of the secular theory are defined through the momenta conjugated
to the mean longitudes and arguments of periapse $g_j=\omega_j$:
\begin{equation}
\begin{array}{l}
\displaystyle
L_{\idm{b}} = \frac{m_{\star} m_{\idm{b}}}{m_{\star} + m_{\idm{b}}}
              \sqrt{G (m_{\star} + m_{\idm{b}})  a_{\idm{b}} },	 \\[0.4cm]
\displaystyle	      
L_{\idm{c}} = \frac{(m_{\star}+m_{\idm{b}}) m_{\idm{c}}}{m_{\star} + m_{\idm{b}}+ 
              m_{\idm{c}}}
              \sqrt{G (m_{\star} +m_{\idm{b}} +m_{\idm{c}}) a_{\idm{c}} },\\[0.4cm]
\displaystyle	      	      
 G_j = L_j \sqrt{1-e_j^2},
\end{array}
\label{eq:forme}
\end{equation}
where $m_{\star}$ is the mass of the parent star. The momenta are,
respectively, the magnitude of the maximum possible angular momentum (for
circular orbits) and the magnitude of the angular momentum. The sum of the 
remaining two momenta, $H_j = G_j \cos i_j$, is the $z$-component of the
angular momentum of the system conjugate to the longitudes of the ascending
node $h_j=\Omega_j$. Let us note, that this formulation is valid in the
general case of an inclined system. In coplanar configurations $\omega_j$ are
undefined but then the longitudes of periastron $\varpi_j$ take over their
role. 
In the averaged system, $l_j$ are absent in the Hamiltonian and $L_j$
are constants of motion. Because the octopole-level Hamiltonian depends only
on the combination $\varpi_{\idm{b}}-\varpi_{\idm{c}}$, it turns out that the
total angular momentum  $G_{\idm{b}}+G_{\idm{c}}$  is also the integral of
motion in the averaged system. This reduces the coplanar dynamics to 1 degree
of freedom, with $e_{\idm{b}}$ or $e_{\idm{c}}$, and $\theta$ as the
relevant phase-space variables.  
The authors found that for the same constants:
\[
 \beta = \frac{5}{4} 
 	\frac{(m_{\star} - m_{\idm{b}})}{(m_{\star}+m_{\idm{b}})} \alpha,
\]
$\lambda=L_{\idm{b}}/L_{\idm{c}}$,
the same initial $e_{\idm{b}}, e_{\idm{c}}$ and
$\theta=\varpi_{\idm{b}}-\varpi_{\idm{c}}$ the averaged equations of motion
describe the same trajectories in the phase-space diagram of
$e_{\idm{b,c}}(\theta)$. For planetary masses much smaller than the
central body, these trajectories and the amplitudes of 
eccentricity oscillations should be independent of the inclination
of the system. Further, by introducing the parameter
$
\gamma = (G_{\idm{b}} + G_{\idm{c}})/(L_{\idm{b}}+L_{\idm{c}})
$       
which describes the non-dimensional total angular momentum of the system,
one defines the critical value of $\lambda$, 
\[
\lambda_{\idm{crit}} = \frac{2\gamma^2}{5-3\gamma^2}.
\]
According to the equations of motion in the averaged system,
if $\lambda \simeq \lambda_{\idm{crit}}$ then libration of $\theta$
is almost certain with possibly large amplitude variations of both
eccentricities. In general, the librations can
take place about two libration centers, $0^{\deg}$ or $180^{\deg}$.
Using the fits published in \cite{Fischer2003}, the  authors found that
indeed, for \pstaro system,  $\lambda \simeq \lambda_{\idm{crit}}$ and they
identified an extended  libration island of the SAR with the apsides antialigned
in the exact resonance.

Following \cite{Lee2003}, we examined the ICs for the \pstar and \pstaro
systems by calculating these phase-space diagrams numerically. To obtain
such diagrams, we fixed the constant level of the total angular momentum 
corresponding to the nominal IC and the phase curves were computed for varied
$e_{\idm{c}}$ ($e_{\idm{b}}$ was calculated from the total
momentum integral, the formula~\ref{eq:forme}).  The time of integrations was
the same as  for the derivation of the 2-dimensional stability maps.
The results are shown in Figure~\ref{fig:fig11}.  For both ICI of the \pstar
and IC0 of the \pstaro systems, the
phase space is occupied by  two large libration islands about
$\theta=0^{\circ}$ and $\theta=180^{\circ}$ which is consistent with the
secular theory. For the ICI of the \pstar system, $\lambda=0.644$ and 
$\lambda_{\idm{crit}}=0.9$, so these parameters substantially differ.
However, Fig.~\ref{fig:fig11} shows that  the occurrence of a libration
mode still appears more likely  than the $\theta$ rotations. For the
IC0 of the \pstaro, $\lambda=0.75$ is closer to its  critical value
$\lambda_{\idm{crit}}=0.87$, and indeed, the the resonance island is more
extended.

The configuration corresponding to ICI of the \pstar system  
is localized almost on the border of
the island of the antialigned SAR. This border is a separatrix and if the
initial $e_{\idm{c}} \simeq 0$, the semi-amplitude of $\theta$ librations
approaches $90^{\circ}$ as the system enters the vicinity of the separatrix. 
In the resonance island, the librations are about the libration center
$\theta=180^{\circ}$.  The
diagram also helps us to understand the large amplitudes of the
eccentricities. Qualitatively, one obtains the same picture  for the
\pstaro system.
For the ICII of the \pstar system, we have $\lambda=0.97$ and $\lambda_{\idm{crit}}=0.77$. 
In this case the nominal  \pstar system lies in an
extended area of $\theta$-rotations seen in Fig.~\ref{fig:fig11}.

The octupole theory allows us to quickly examine the extent of the SAR
in the space of the orbital parameters. Because the elements of the inner planet
are well determined, the overall picture of the \pstar dynamics depends
mostly on the not too well constrained parameters of the outer planet. Assuming that
strong mean motion resonances are absent, the secular dynamics can be
investigated in detail through solving Equations (1)---(4) of
 \cite{Lee2003}. These equations describe the secular
time-evolution of $e_{\idm{b,c}}$ and $\varpi_{\idm{b,c}}$.  Unfortunately,
there seems to be no explicit, analytical solution to these equations. Also a
construction of a simple SAR criterion, like the one developed by
\cite{Laughlin2002} or \cite{Ji2003} in the framework of the Laplace-Lagrange
secular theory, does not appear to be possible, either.  However, these
equations are very simple for a numerical treatment and their direct
integration is rapid, hundreds times faster than the $N$-body
integrations. The SAR can be detected in the parameter space by
looking for the maximal deviations of $\theta$ from the given libration
center $0^{\circ}$ or $180^{\circ}$. It is the same method of
detecting the SAR as the one applied in the full $N$-body integrations
but here it is well justified because according to the generic secular
dynamics, the SAR appears about two libration centers. Assuming that the
integration time covers at least one secular period (so as it is not too short), we
can easily detect the librations or rotations of the critical argument
$\theta$.

Using this approach, we computed a number of the  SAR-maps in
the $(a_{\idm{c}}, e_{\idm{c}})$ 
and $(e_{\idm{b}},e_{\idm{c}})$ domains for the
initial conditions ICI, ICII and IC0. In order to directly compare the
results, the ranges of the parameters were the same as for the numerical maps.  
The results are shown in Fig.~\ref{fig:fig12}. Generally, they agree in
the regions corresponding to small eccentricities, for which  both the shape
of these regions 
and the semi-amplitude of librations are well reproduced by the secular
theory. However, we can also find significant discrepancies between the analytical and 
numerical estimations. For example for ICII, 
in the $(a_{\idm{c}},e_{\idm{c}})$-plane 
one can find islands of the SAR which are absent in the map computed by
solving the secular equations. The differences are also apparent in the 
$(e_{\idm{b}},e_{\idm{c}})$ plane for all examined IC's. In these cases, 
the results agree only in the regions of relatively small $e_{\idm{c}}$. 

In order to understand these differences, we compared the 
$(e_{\idm{c}},\theta)$ diagrams obtained  by the direct numerical integrations 
and by solving the secular equations. The tests were performed for: (a) ICI
in which we changed eccentricities to $e_{\idm{b}}=0.3,e_{\idm{b}}=0.3$, (b) 
ICII with $a_{\idm{c}}=3.51,e_{\idm{c}}=0.47$ (see bottom-middle panel in
Fig.~\ref{fig:fig7} corresponding to the center of the 9:1 MMR) (c) ICII with
$(e_{\idm{b}}=0.47,e_{\idm{b}}=0.47)$ (corresponding to the island in the
bottom middle panel in Fig.~\ref{fig:fig8}). The results are presented in
Fig.~\ref{fig:fig13}. In all these cases, the full dynamics seems to be
only roughly described by the secular approximation. The disagreement is
obviously expected in the case (b) since the assumptions  of the secular
theory are violated (an MMR is present). In other two examples, the 
differences are most likely due to significant mutual interactions caused
by large eccentricities resulting in deformations of
the phase-space curves (case a) or a chaotic evolution
(case c). Clearly the secular theory should be applied with caution.

\section{Conclusions}

In this paper we carry out a dynamical analysis of the recently discovered
2-planetary exosystem around \pstaro. According to the preliminary initial
orbital parameters announced by the Geneva Extrasolar Planet Search
team (orbital solution ICI), this system resembled the other known planetary 
hierarchical exosystem HD~12661 and the HD~160691. The recently updated
orbital solution (ICII) has been changed in a significant way and the 
close similarity of the \pstar and \pstaro systems is not so apparent
anymore.  

The exosystems around HD~122661, HD~160691 and \pstar belong to a class of
the hierarchical planetary systems. They are characterized by
a relatively small ratio of the semi-major axes, $\simeq 0.1-0.3$. 
Their dynamics should be analyzed in the Jacobi orbital elements 
\citep{Lee2003,Gozdziewski2003e}. The commonly used astrocentric elements
may  lead to small shifts of the positions of the orbital resonances and to
differences  in the evolution of the osculating orbital elements.  Following
these papers we used the two orbital fits published by the discovery team but
we recomputed the semi-major axes and masses of the companions. The current
ICII was verified by our independent analysis of the approximate RV
measurements which we obtained by digitizing a figure with
the synthetic best-ft RV curve and the real observational points. Such an "approach" to
obtain the "observations" was forced by the lack of access to the real data.
Since the $\Chi$ function will typically have many local minima, we cannot be 
sure that a solution is the proper one  without a global $\Chi$ 
analysis. As demonstrated on the solution ICI and ICII of the \pstar system,
the difference between the global and a local minimum may lead to enormous differences 
in the overall dynamical picture of the system. Obviously, the digitization cannot provide
very accurate moments of the observations, nevertheless such "data" describe
the the overall shape of the real RV measurements quite precisely and make it
possible to obtain some insight into the quality of the best fit solution by
the discovery team. Using the genetic algorithm to find the global minimum 
of $\Chi$ and describing the measurements with the 2-planet Keplerian and Newtonian
models, we  found almost the same orbital
elements of the inner planet as these reported by the discovery team and very
similar elements for the outer companion. The application of the 2-Keplerian
and $N$-body models produced the same value of $\Chi$ and similar orbital solutions.
It demonstrates the absence of strong interactions between the
planets. It also favors the solution with a large separation between the
companions rather than the previously announced configuration. Apparently, 
the first orbital solution was just a local minimum of the $\Chi$ function.

The two orbital solutions have different dynamical features. The ICI
solution is close to an unstable 13:2 MMR and lies between 6:1 and 7:1 MMRs.
The exosystem corresponding to ICII is located between 9:1 and 10:1 MMRs.
Both configurations appear to be stable on the Gyr time scale as demonstrated by 
our long-term integrations, the fast indicator analysis and the secular octupole-level 
theory. In the phase space, both ICs are lie in the wide regions of stable motions
in spite of large variations of the eccentricities (up to 0.5). Obviously, these 
conclusions may changed in the future as the time-span of the current RV measurements 
is shorter than the period of the outer planet. Moreover, the moderately constrained 
orbital elements of the outer planet do not allow us to exclude a proximity of the 
system to an unstable low-order resonance. The main dynamical difference between the 
two ICs is the SAR with the apsidal lines antialigned in the exact resonance. 
It is present for ICI in the wide ranges of the orbital parameters while for the 
ICII it seems to be ruled out. In the first case, the \pstar would be very similar 
to the hierarchical resonant systems. In this "class" we can find the discussed 
HD~12661 and possibly the HD~160691 exosystem. The HD~37124~\citep{Butler2003} 
system could also be a member of this group. However, in the HD~37124 system the 
apsides are most likely aligned \citep{Ji2003}. If the two outer planets are taken 
into consideration, the $\upsilon$~Andromedae~\citep{Butler1999} system could also
be assigned to this group \citep{Chiang2002}. However, the new orbital solution ICII
favors the proximity of the \pstar system to the HPS with the outer, very massive 
(possibly sub-stellar) companion like the HD~38529~\citep{Fischer2003}, 
HD~74156\footnote{\url{http://obswww.unige.ch/~udry/planet/hd74156.html}} and
HD~168443~\citep{Marcy2001,Udry2002} systems. In these systems, the stability is
maintained thanks to a relatively large separation of the companions.
Nevertheless, due to very large masses involved and large amplitudes of
the eccentricity variations, the mechanism providing the stability is
puzzling. 

The results of the numerical analysis of the orbital evolution and  stability
can be compared with the predictions by the secular
octupole-level theory of \cite{Lee2003}. The
results of both approaches are in accord for moderately low eccentricities of the outer planet.
The discrepancies appear if both eccentricities are large in the regions of
the phase space close to the MMRs and in the chaotic zones. It seems that
although the secular theory makes it possible to explain the main features of
the \pstar-like systems, the direct integrations are still necessary to
understand the dynamics in detail.

For a better grasp on the dynamical picture of the \pstar system, 
we need to wait for some time, ideally until the observations cover a few 
orbital periods of the outer companion. Since the orbital elements of the inner
companions are similar in the \pstar and \pstaro systems and are 
already determined with high accuracy, there is still a chance that our
preliminary hypothesis that the two planetary systems are dynamically
related is plausible.

\section{Acknowledgments}
We thank Man Hoi Lee for a discussion about the Jacobi elements
and  Ji Janghui for useful remarks.
 This work is supported by the
Polish Committee for Scientific Research, Grant No.~2P03D~001~22. M.~K. is a
Michelson Postdoctoral Fellow.

\bibliographystyle{apj}
\bibliography{ms}

\clearpage

%
%

\figcaption[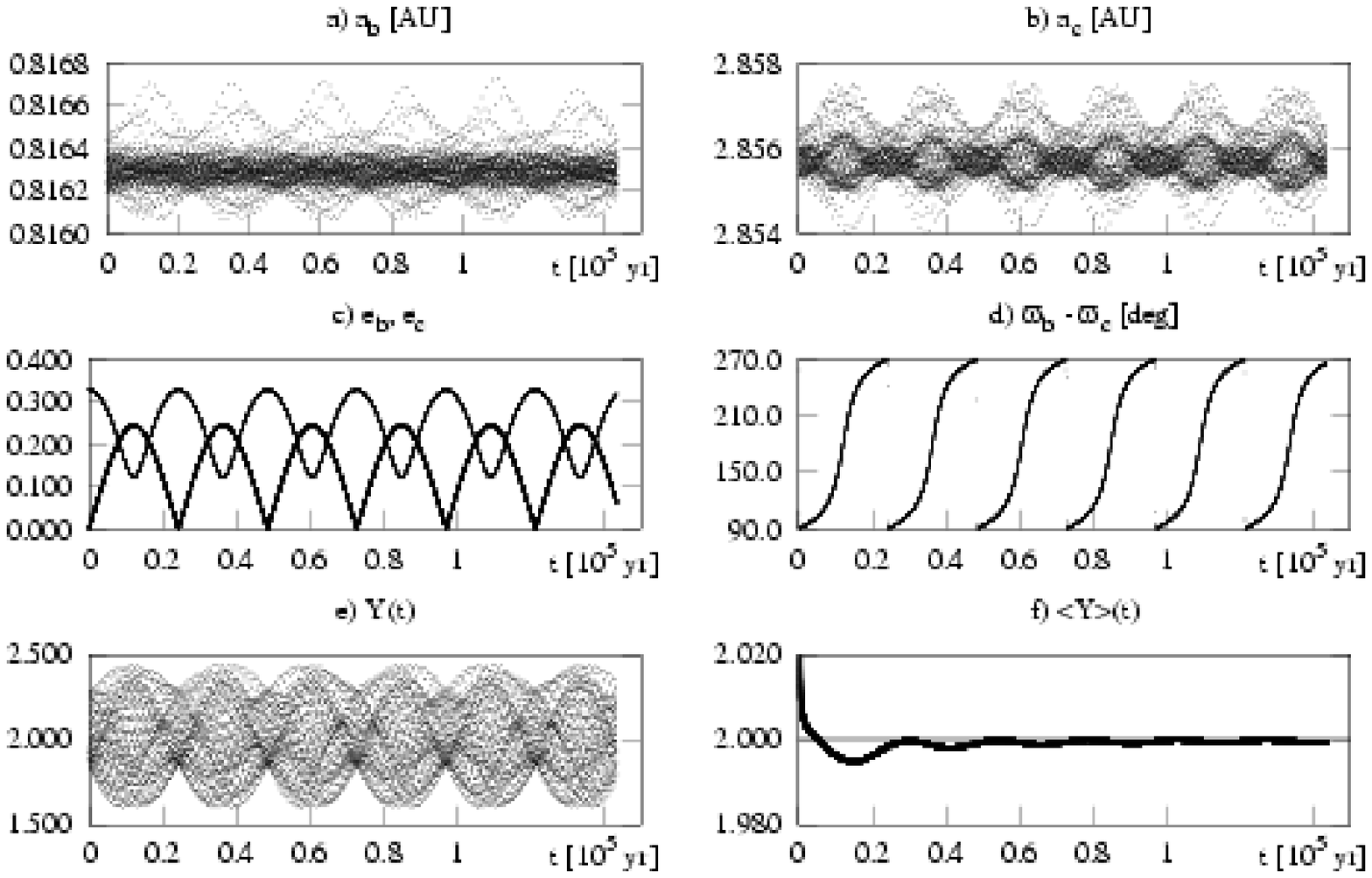]{
The orbital evolution (expressed in the osculating orbital elements of
Jacobi)  of the \pstar system for the nominal ICI given in Table
\ref{tab:tab1}. Panels ({\bf a,b}) show changes of the semi-major axes. Panel
({\bf c}) is for the eccentricities. Panel ({\bf d}) shows the critical
argument of the apsidal resonance $\theta = \varpi_{\idm{b}} -
\varpi_{\idm{c}}$.  Panel ({\bf e}) is for MEGNO as a function of time,
$Y(t)$, and  panel ({\bf f}) is  for its mean value $\Ym$.}

\figcaption[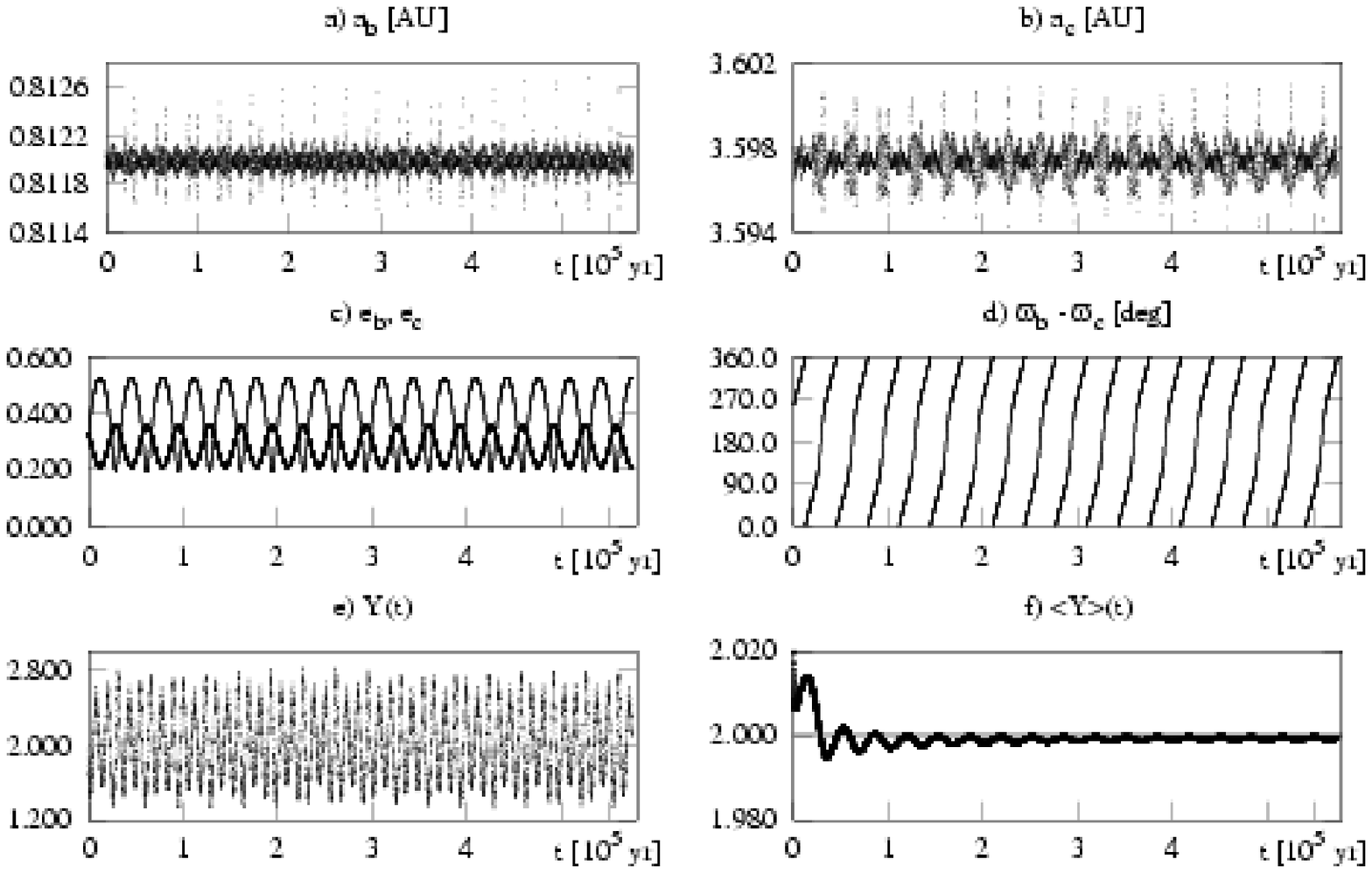]{The orbital evolution of the \pstar planetary system for
the ICII given in Table \ref{tab:tab1}. The description of the panels is the
same as in the previous figure.}

\figcaption[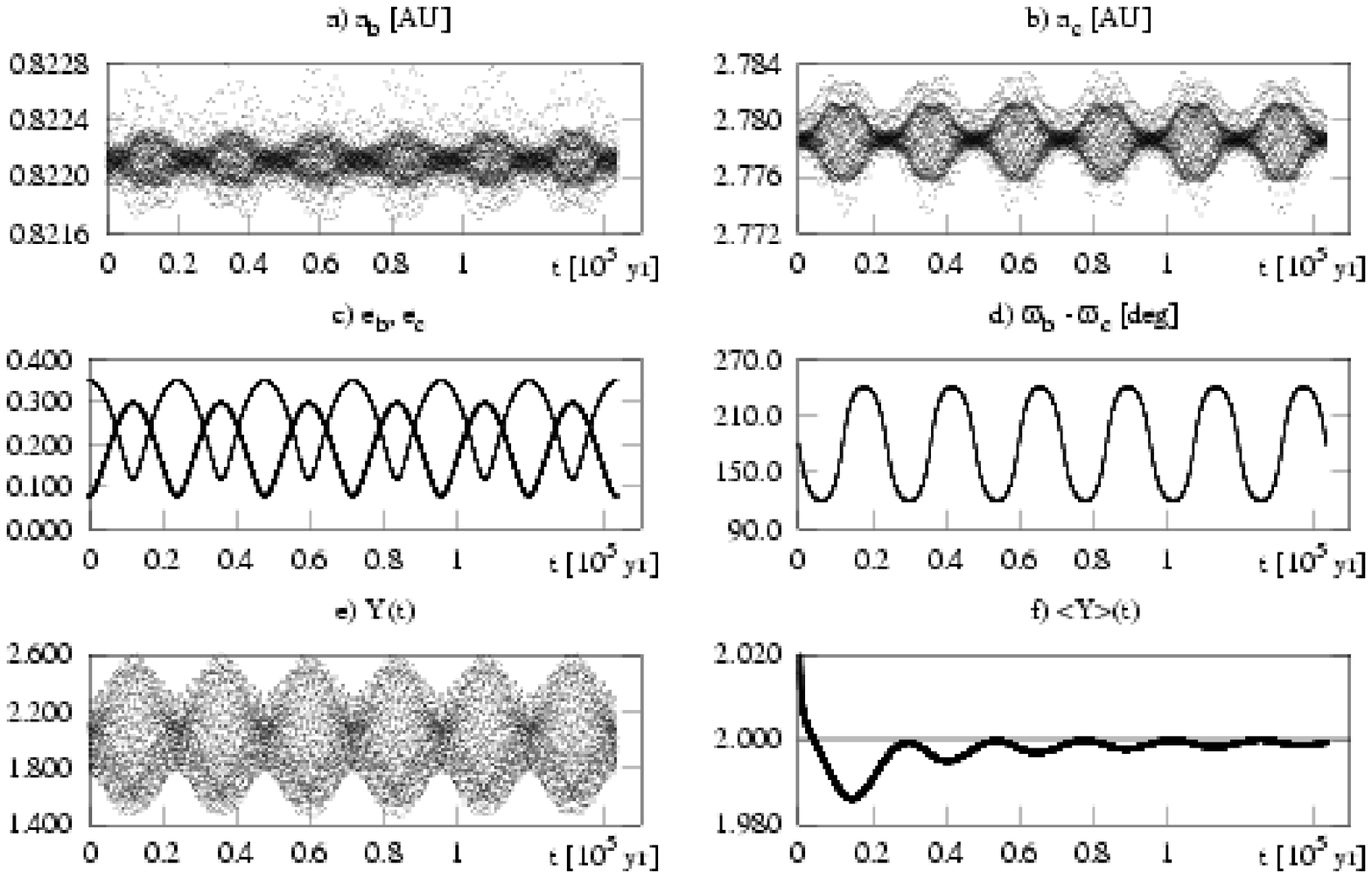]{The orbital evolution of the HD12661 planetary system for
the IC given in Table \ref{tab:tab3}. The description of the panels is the
same as in the previous figure.}

\figcaption[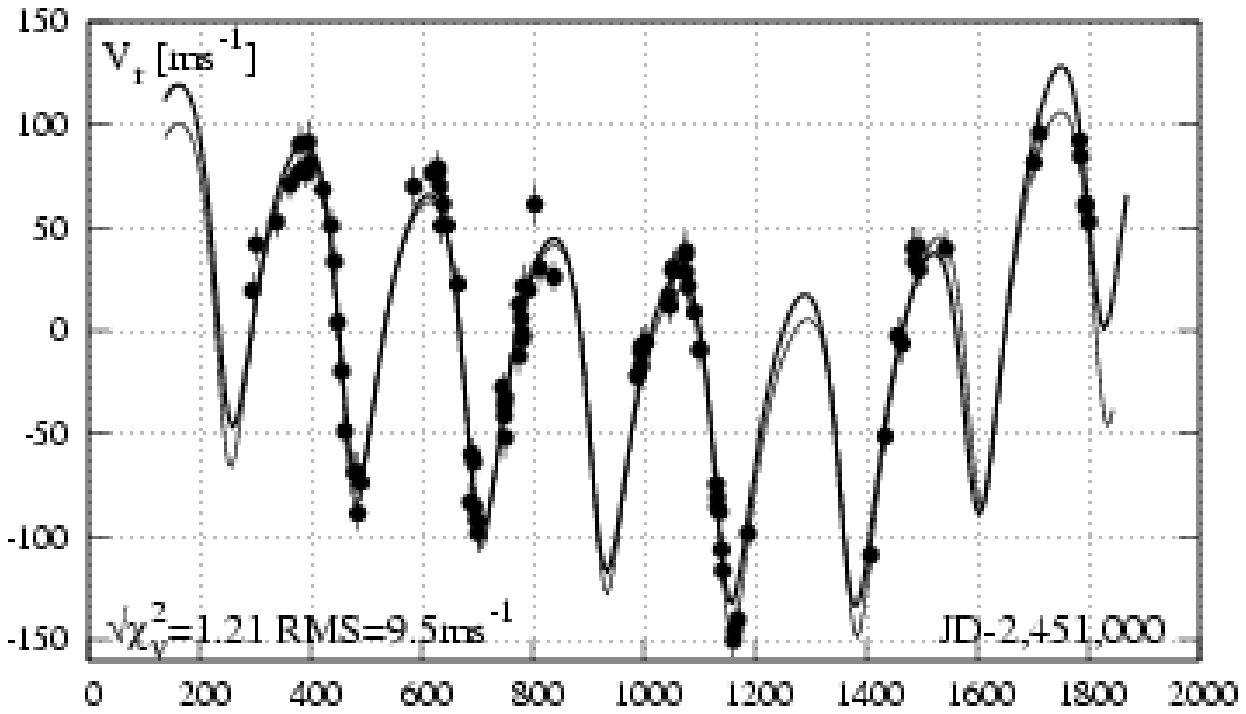]{
The digitized RV data of the
\pstar system and the synthetic RV curves corresponding
to the best fits found from this paper. The thicker line is for the
$N$-body model. The orbital parameters of the solutions are given in
Table \ref{tab:tab2}. 
}

\figcaption[f5a.eps,f5b.eps,f5c.eps]{
Confidence intervals corresponding to
$1\sigma$, $2\sigma$ and $3\sigma$ levels of $\Chi$ obtained for
the best $N$-body fit to the digitized RV data. The fit
is given in Table \ref{tab:tab2}. 
}

\figcaption[f6a.eps,f6b.eps,f6c.eps]{
Scans of MEGNO (upper graphs in every panel)  and the diffusion rate
of the proper mean motion of the
outer planet (lower graphs labeled by $\log \sigma_{\idm{c}}$) for the
ICI of  the \pstar system (the top panel) 
ICII of  the \pstar system (the middle panel)
and for the
HD~12661 system (the bottom panel) over the semi-major axis of the outer
planet (in AU).  
The resolution of the plot is $\simeq 3 \cdot 10^{-4}$~AU. Labels
mark the positions of the strongest  mean motion resonances (up to the order
about 20) as they have been identified by the FMA. 
The nominal values of $a_{\idm{c}}$ are marked with big dots.}

\figcaption[f7a.eps,f7b.eps,f7c.eps,f7d.eps,f7e.eps,f7f.eps]{
The stability of the \pstar system in the $(a_{\idm{c}},e_{\idm{c}})$-plane.
The left panels are for MEGNO. The middle panels give an estimate of the
semi-amplitude of the apsidal librations, $\theta^{\idm{max}}$, about the
line of apsides antialignment (note
that the white areas are for $\theta^{\idm{max}} \geq
90^{\deg}$).
The right panels show the maximal
eccentricity of the inner planet.  The data grid has the resolution of
$0.006\mbox{AU} \times 0.005$.
The top row is for the ICI, the bottom row is for the ICII.
}

\figcaption[f8a.eps,f8b.eps,f8c.eps,f8d.eps,f8e.eps,f8f.eps]{
The stability of the \pstar system in the plane of the eccentricities.  The
left panels are for MEGNO. The middle panels give an estimate of the
semi-amplitude of the apsidal librations, $\theta^{\idm{max}}$, about
$180^{\circ}$. Note that the white areas are for $\theta^{\idm{max}} \geq
90^{\deg}$. The right panels show the maximal eccentricity of the inner
planet. The data grid has the resolution of $100 \times 100$ points.
The top row is for the ICI, the bottom row is for the ICII.
}

\figcaption[f9a.eps,f9b.eps,f9c.eps,f9d.eps,f9e.eps,f9f.eps]{
The stability map of the \pstar system when  the relative inclination of the
orbits and the absolute inclination $i$ are varied. The left panels are for
MEGNO. The middle panels show the semi-amplitude of librations of
$\theta$  about $180^{\circ}$. 
The white areas are for $\theta^{\idm{max}} \geq
90^{\deg}$. The right panels show the maximal
eccentricity of the inner planet.  The data grid  has the resolution
of $3^{\circ}\times 1^{\circ}$. The top row is for the ICI, the bottom row 
is for the ICII.
}

\figcaption[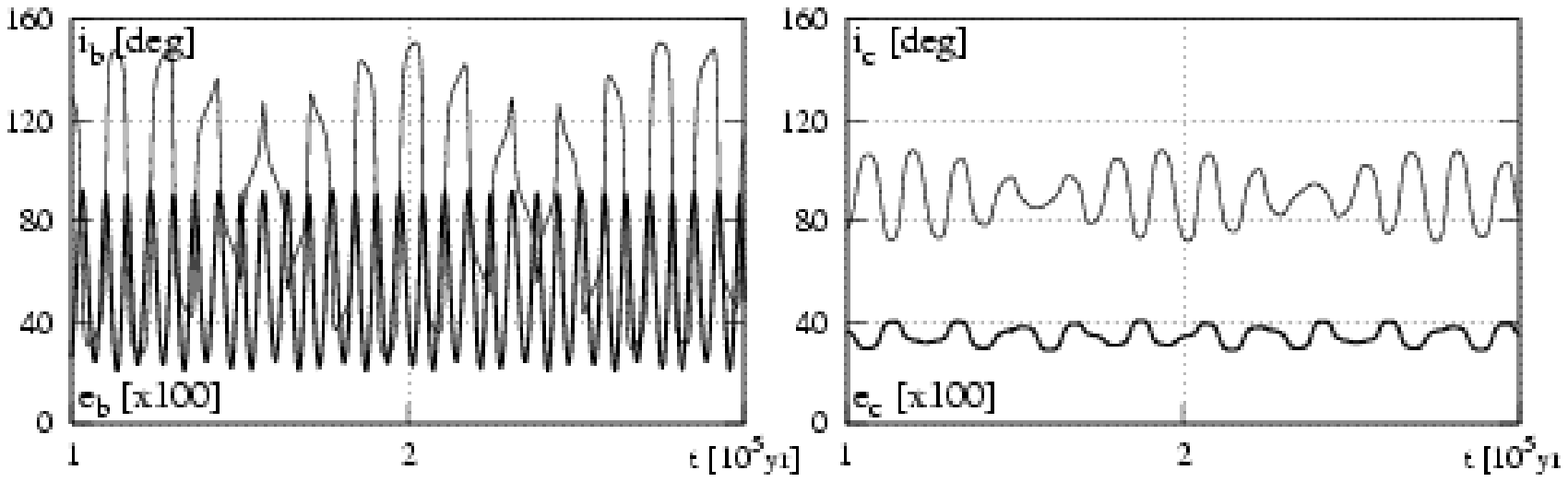]{
The evolution of the inclination and eccentricities
in the planetary configuration corresponding to ICII
when the initial relative inclination of orbits has been changed
to $80^{\circ}$. The left panel is for the inner planet and the
right panel is for the outer companion.
The bottom plots are for the eccentricities
magnified by 100. 
}

\figcaption[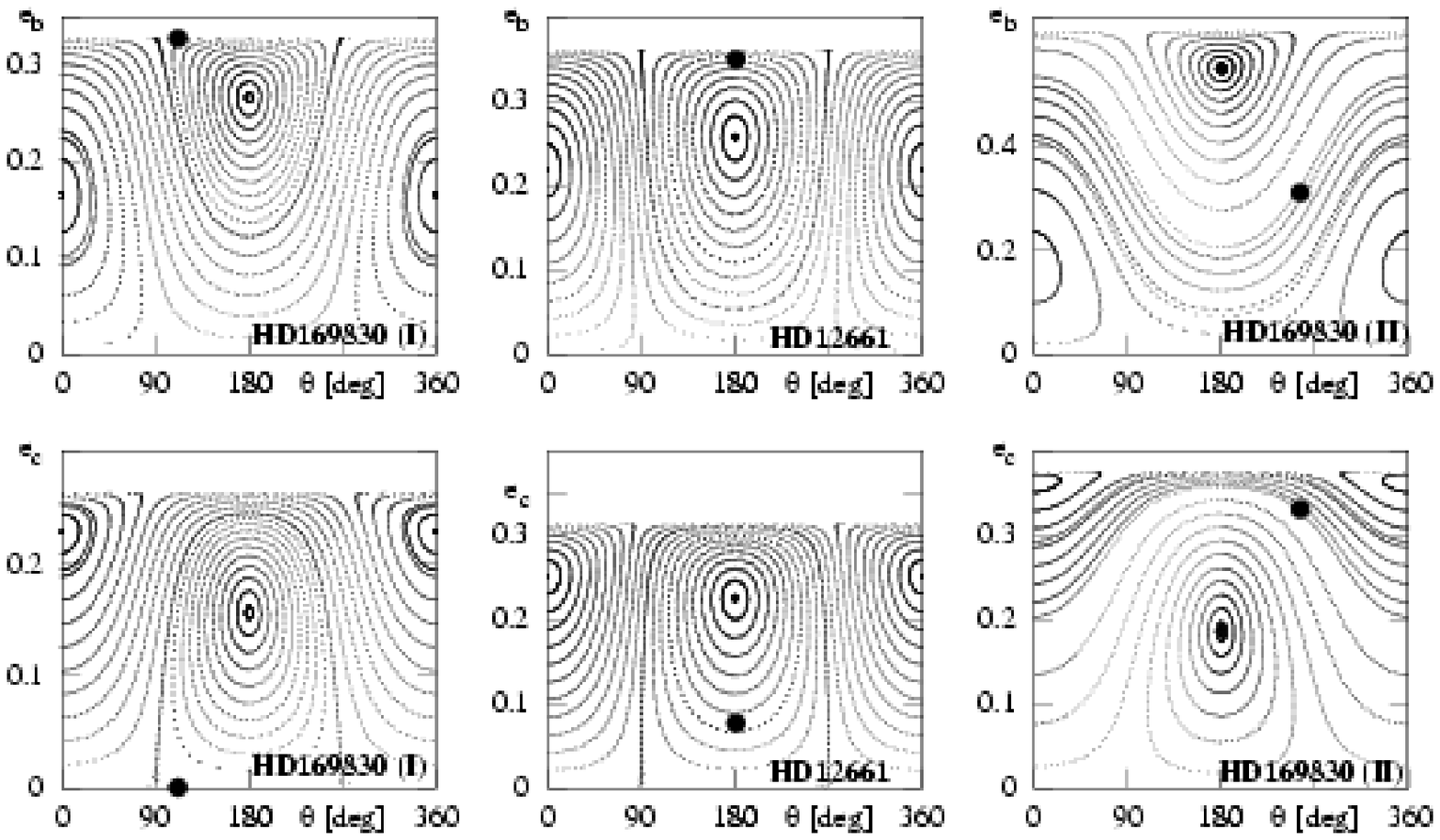]{
The phase-space diagrams at the same level of the total angular  momentum.
The left column is for the ICI, the middle column is  is for the HD~12661
system, the right column is for the ICII. The nominal ICs are marked with a
big dot. The phase-space trajectories have been computed for the initial
$\theta=0^{\circ}$ and for $\theta=180^{\circ}$; $e_{\idm{c}}$ varies while
$e_{\idm{b}}$ is determined from the integral of the total angular momentum.
Other orbital parameters are fixed at their nominal values.
}

\figcaption[f12a.eps,f12b.eps,f12c.eps,f12d.eps,f12e.eps,f12f.eps]{
The semi-amplitude of the SAR about the libration center $180^{\circ}$
in the planes of semi-major axes and eccentricities as predicted
by the secular octupole theory. The left column is for the ICI,
the middle column is for the ICII of the \pstar system and 
the right column is for the IC0 of the \pstaro system.
The white areas are for $\theta^{\idm{max}} \geq
90^{\deg}$.
}

\figcaption[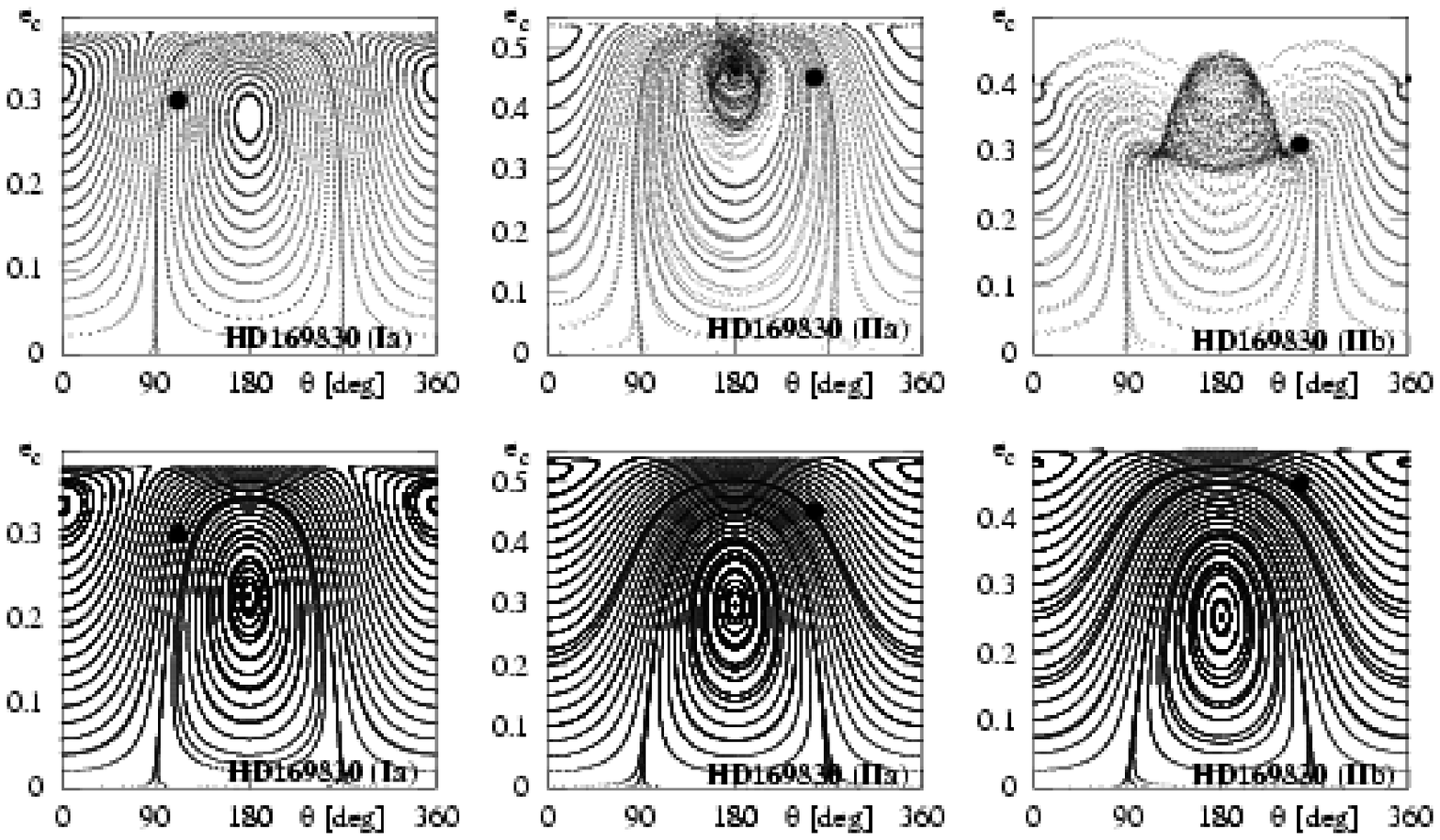]{
The comparison of the  phase-space diagrams at the same level of the  total
angular momentum obtained from the numerical integration of the full
equations of motion (top row) and derived from the secular-octupole
theory (bottom row). The initial conditions correspond to the modified ICI
and ICII (see
text for explanation). The modified initial conditions are marked with
big dots. They correspond to the coordinates marked with (a), (b) and
(c) in Fig.~\ref{fig:fig7}, \ref{fig:fig8}, respectively.
}

\clearpage

%
%

\begin{figure}
\figurenum{1}
\centering
\hbox{\includegraphics[]{f1.eps}}
\caption{}
\label{fig:fig1}
\end{figure}

%
%

\begin{figure}
\figurenum{2}
\centering
\hbox{\includegraphics[]{f2.eps}}
\caption{}
\label{fig:fig2}
\end{figure}

%
%

\begin{figure}
\figurenum{3}
\centering
\hbox{\includegraphics[]{f3.eps}}
\caption{}
\label{fig:fig3}
\end{figure}

%
%

\begin{figure}
\figurenum{4}
\centering
\hbox{\includegraphics[]{f4.eps}}
\caption{}
\label{fig:fig4}
\end{figure}

%
%

\begin{figure}
\figurenum{5}
\centering
\includegraphics[]{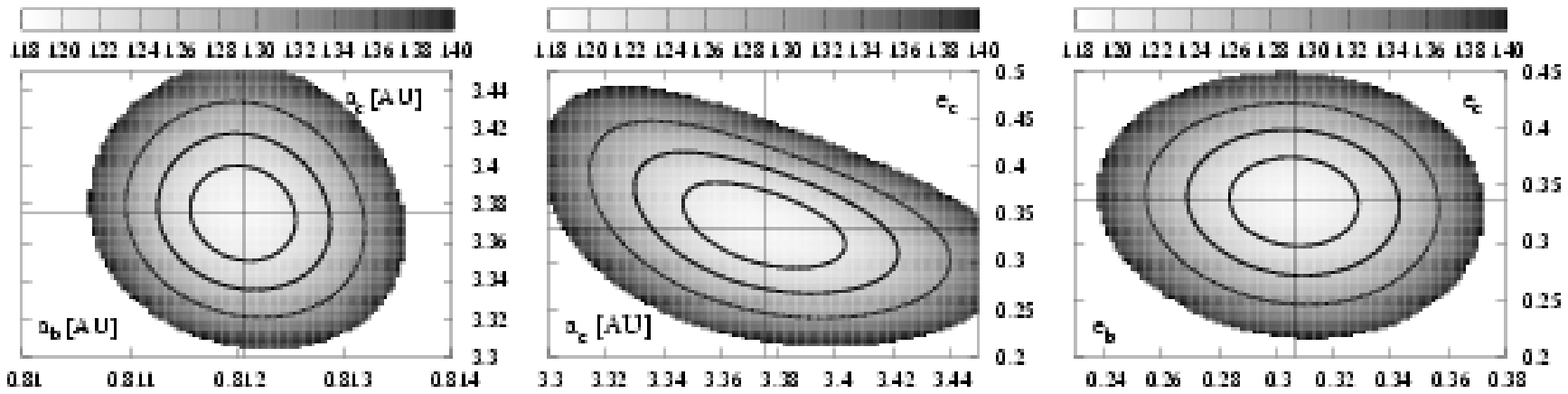}
\caption{}
\label{fig:fig5}
\end{figure}

%
%

\begin{figure}
\figurenum{6}
\centering
\hbox{\includegraphics[]{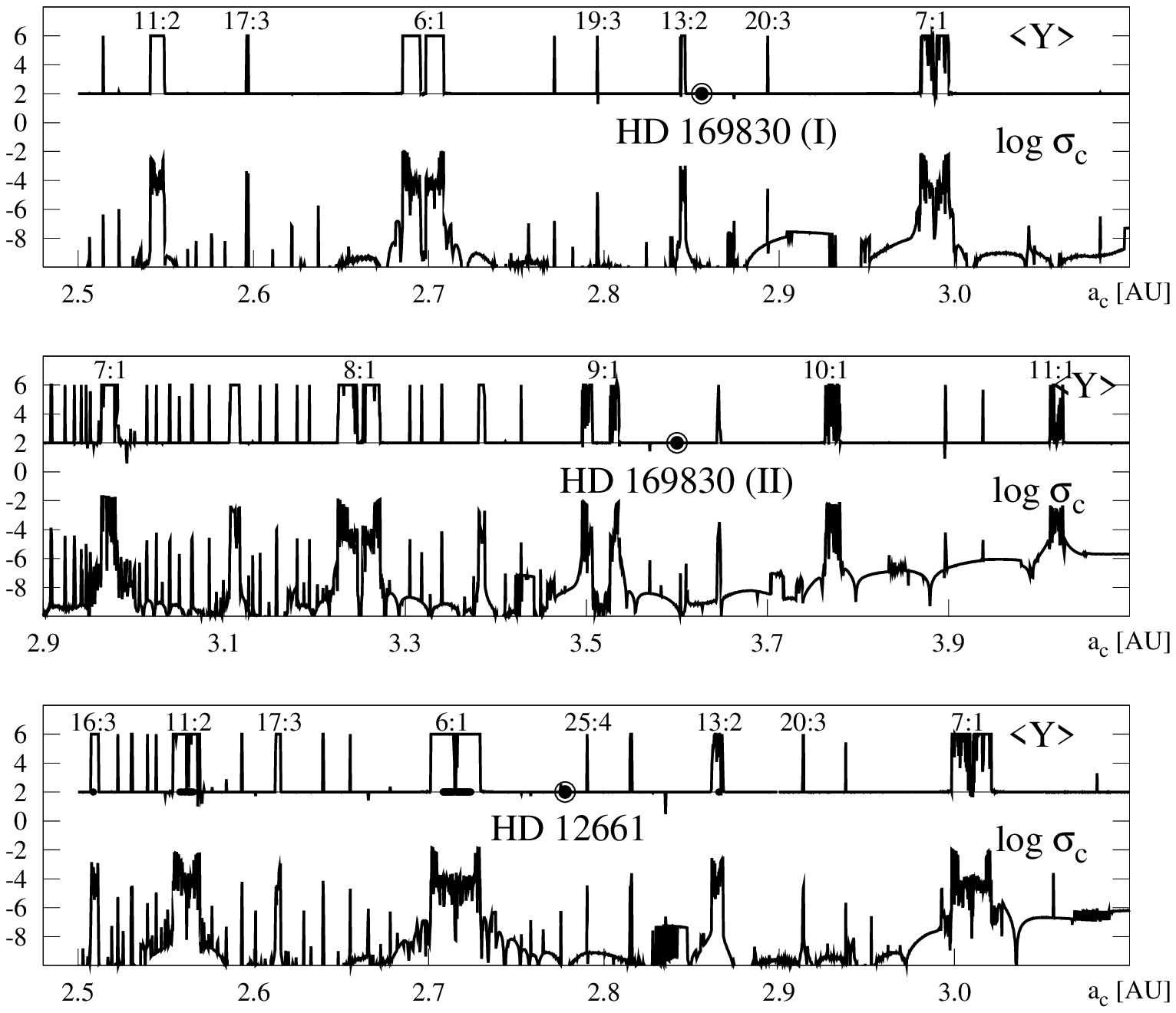}}
\caption{}
\label{fig:fig6}
\end{figure}

%
%

\begin{figure}
\figurenum{7}
\centering
\includegraphics[]{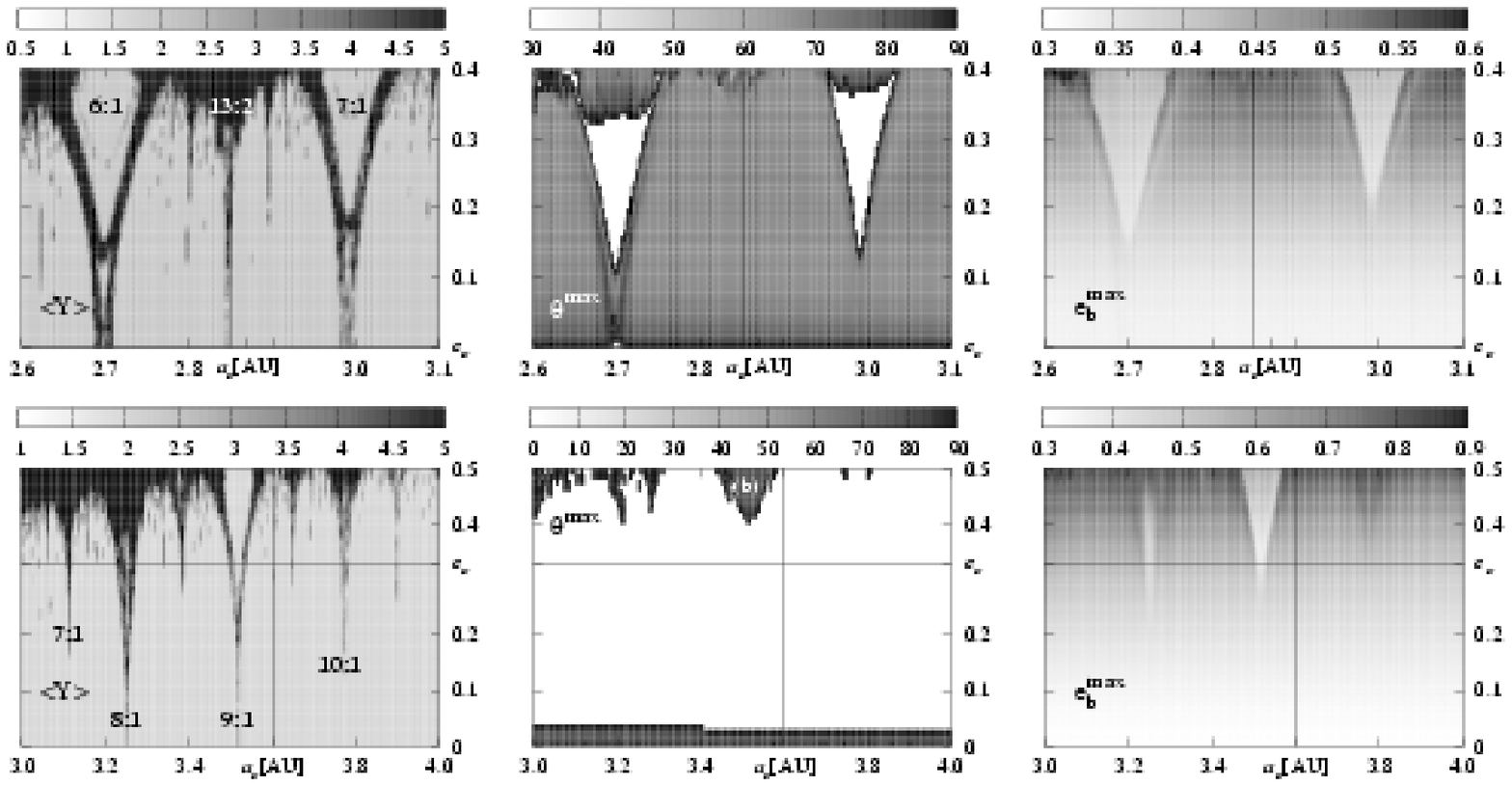}
\caption{}
\label{fig:fig7}
\end{figure}

%
%

\begin{figure}
\figurenum{8}
\centering
\includegraphics[]{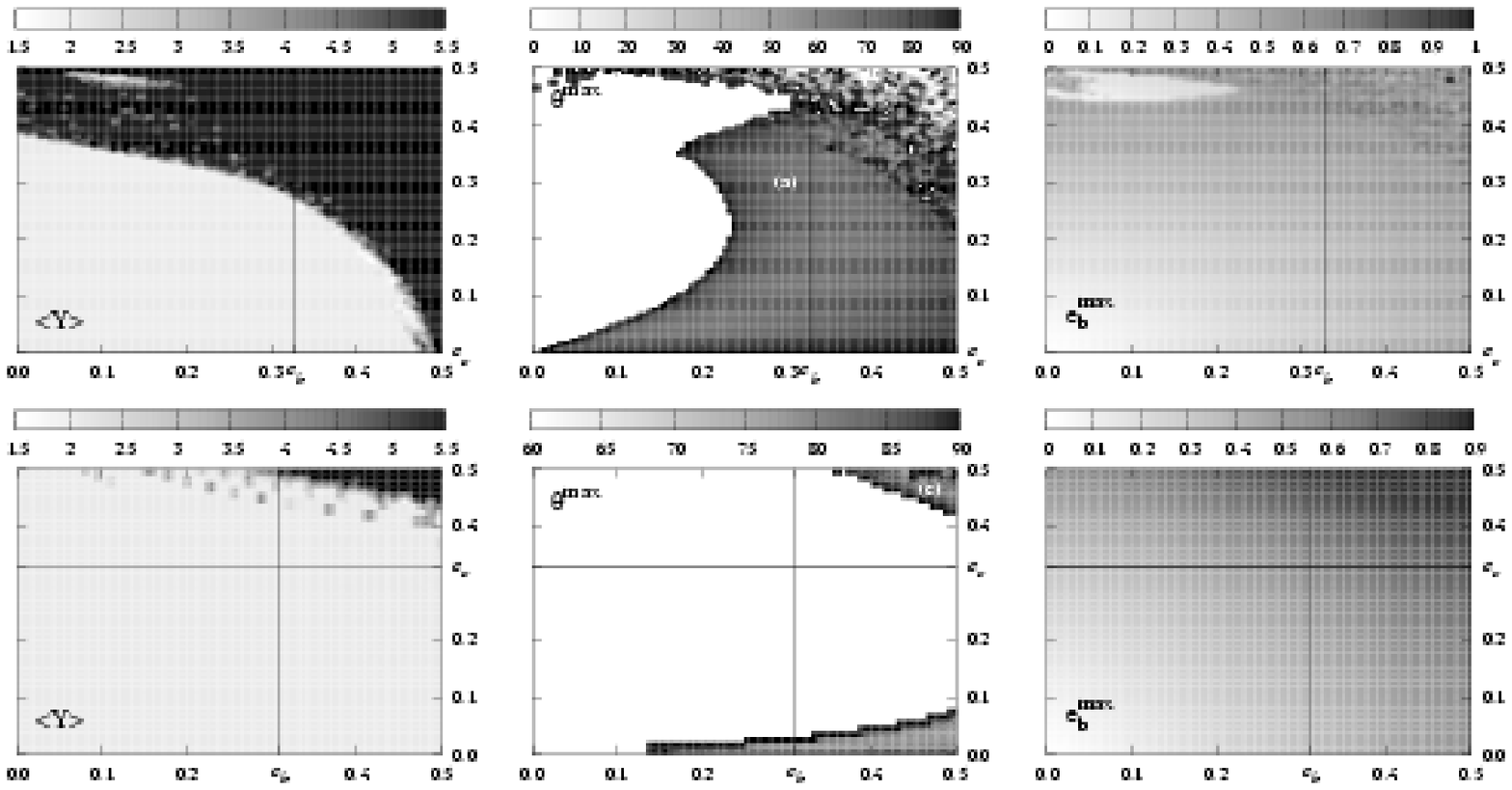}
\caption{}
\label{fig:fig8}
\end{figure}

%
%

\begin{figure}
\figurenum{9}
\includegraphics[]{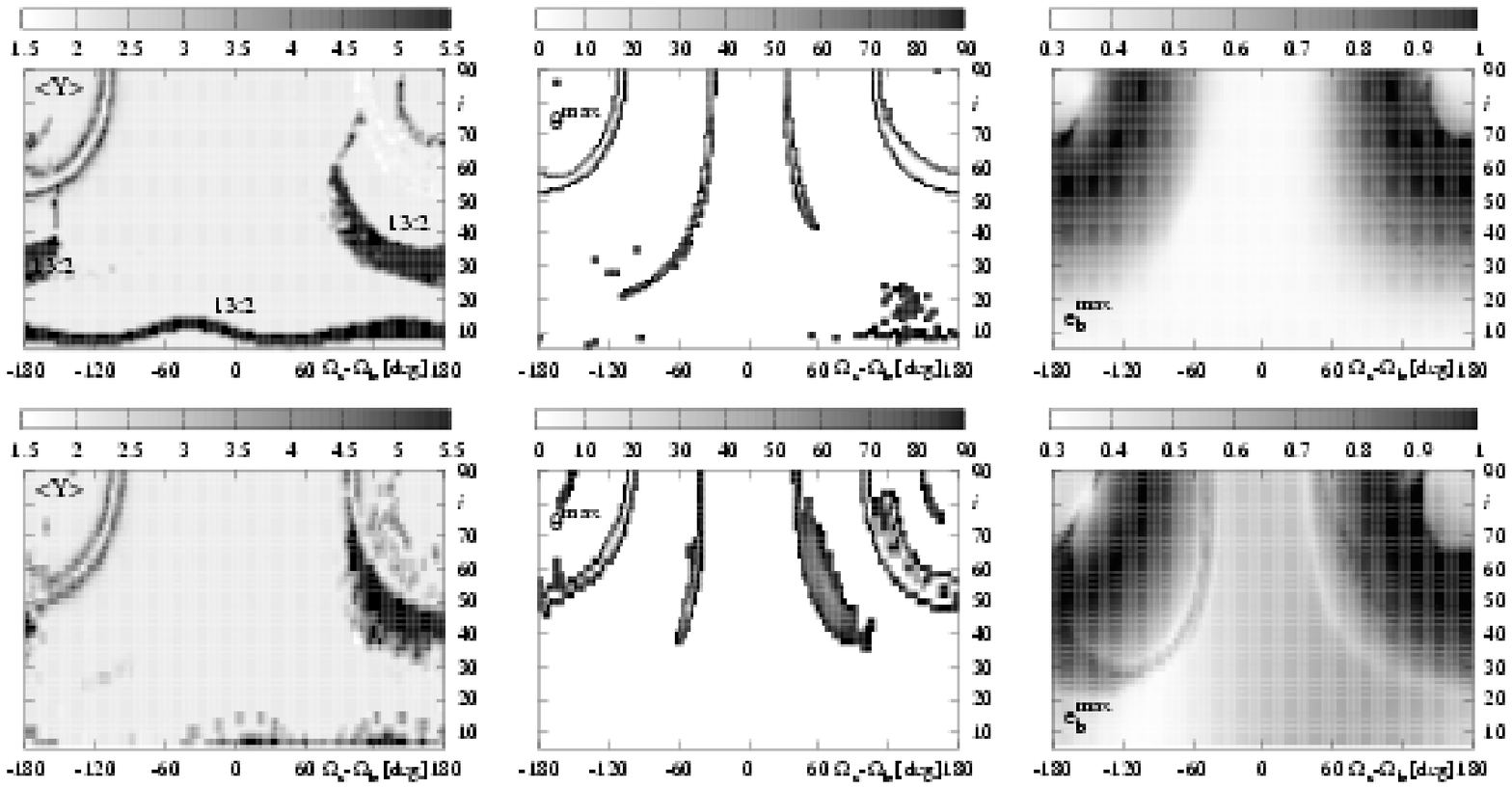}
\caption{}
\label{fig:fig9}
\end{figure}

%
%

\begin{figure}
\figurenum{10}
\centering
\hbox{\includegraphics[]{f10.eps}}
\caption{}
\label{fig:fig10}
\end{figure}

%
%

\begin{figure}
\figurenum{11}
\centering
\hbox{\includegraphics[]{f11.eps}}
\caption{}
\label{fig:fig11}
\end{figure}

%
%

\begin{figure}
\figurenum{12}
\centering
\includegraphics[]{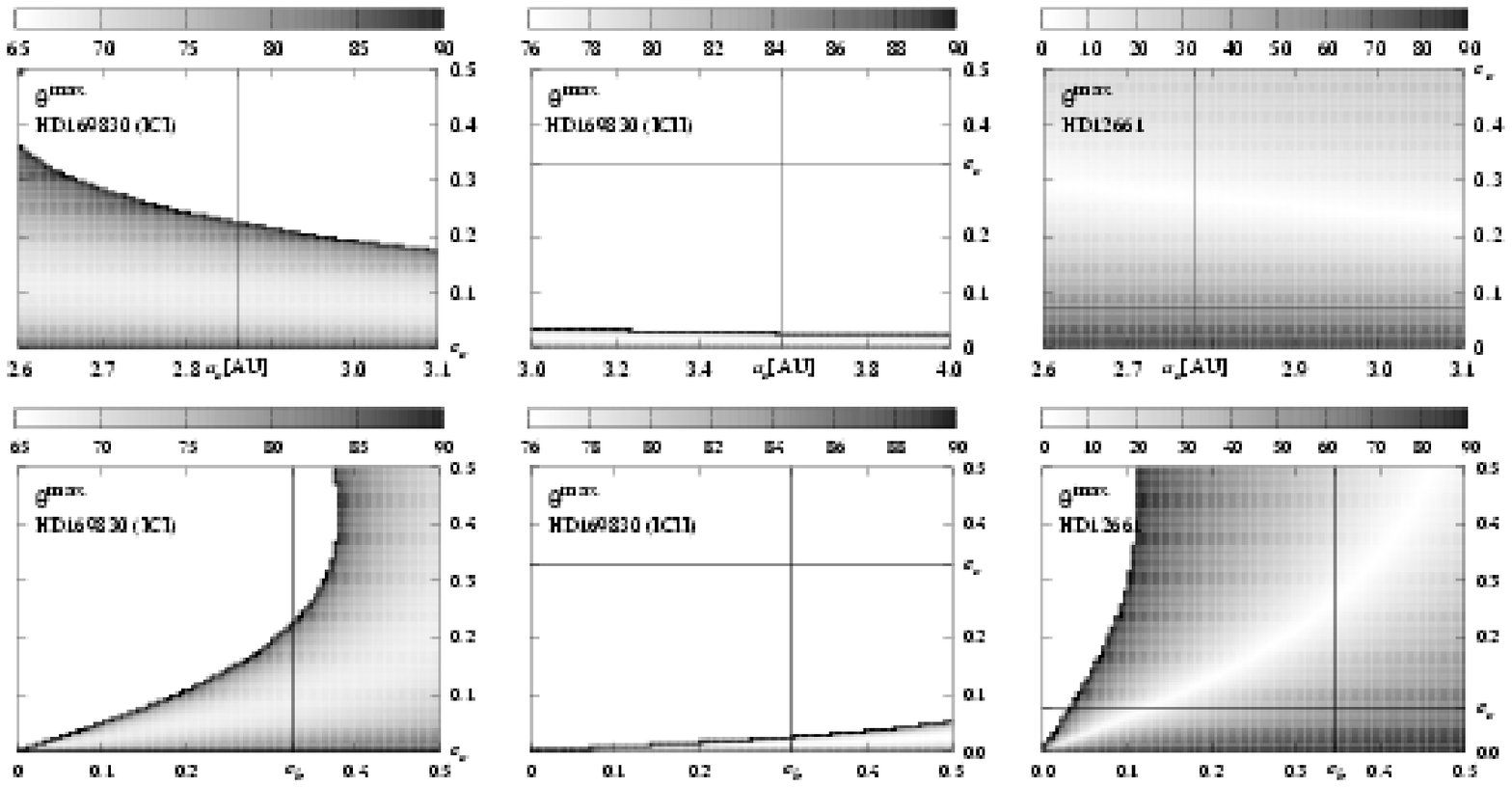}
\caption{}
\label{fig:fig12}
\end{figure}

%
%

\begin{figure}
\figurenum{13}
\centering
\hbox{\includegraphics[]{f13.eps}}
\caption{}
\label{fig:fig13}
\end{figure}

\clearpage

%
%
\begin{table}
\caption{Jacobi
orbital parameters of the \pstar system from
\url{http://obswww.unige.ch/~udry/planet/hd169830\_syst.html}
(updated on August, 28, 2003; the ICII solution).
The first version of the fit (dated 28 June, 2003; the ICI solution) 
is given in the parentheses. The mass of the central star
is equal to $1.4~\mbox{M}_{\sun}$. 
}
\smallskip
\begin{tabular}{lrr}
\hline Jacobi
orbital parameter \hspace{5em}  & \ \ Planet b \ \ & \ \ Planet c \ \ \\
\hline
$\mbox{m}_2 \sin i$ [M$_{\idm{J}}$] \dotfill &  2.88 (3.03) & 4.05 (2.51) \\
a [AU] \dotfill & 0.811 (0.816) & 2.856 (3.598)  \\
P [d] \dotfill &  225.62 $\pm 0.22$ (227.43)   &  2102 $\pm 264$ (1487) \\
e \dotfill     &  0.31 $\pm 0.01$ (0.327) & 0.33 $\pm 0.02$ (0.0) \\
$\omega$ [deg]\dotfill & 148 $\pm 2$ (156.1) & 252 $\pm 8$ (44.0)  \\
$T_{\idm{p}}$ [JD-2,400,000]  \dotfill& 51923 $\pm 1$ (51472.43)  & 52516
$\pm 25$ (50101) \\
$M(T_{\idm{b}})$ [deg] \dotfill & 0.00 & 258.4  (332.2) \\
\hline
\end{tabular}
\label{tab:tab1}
\end{table}

%
%
\begin{table}
\caption{Jacobi
orbital parameters of the \pstar system based on the digitized figure
from \url{http://obswww.unige.ch/~udry/planet/hd169830\_syst.html}
(the version updated on August, 28, 2003). Numbers without parentheses 
are for Jacobi elements of the 2-Keplerian solution.
Values in parentheses are for the osculating, Jacobi
elements from the $N$-body, self-consistent fit given for the
date of the first digitized observation (JD=2,451,294.97).
The mass of the central star
is equal to $1.4~\mbox{M}_{\sun}$. 
}
\smallskip
\begin{tabular}{lrr}
\hline
Orbital parameter \hspace{5em}  & \ \ Planet b \ \ & \ \ Planet c \ \ \\
\hline
$\mbox{m}_2 \sin i$ [M$_{\idm{J}}$] \dotfill &  2.86 (2.85) & 3.88 (3.84) \\
a [AU] \dotfill & 0.812 (0.812) & 3.41 (3.38)  \\
P [d] \dotfill &  225.62    &  1939.4  \\
e \dotfill     &  0.307 (0.307) & 0.334 (0.336) \\
$\omega$ [deg]\dotfill & 148.6 (148.8) & 260.1 (260.5)  \\
$T_{\idm{p}}$ [JD-2,400,000]  \dotfill& 51922.5 & 52533 \\
$M$ [deg] \dotfill & 0.0 (78.9) & 204.7 (128.1) \\
$\Chi$  \dotfill & \multicolumn{2}{c}{1.21 (1.21)} \\
RMS [m/s] \dotfill & \multicolumn{2}{c}{10 (9.5)} \\
$V_0$ [m/s] \dotfill & \multicolumn{2}{c}{3.35 (6.3)} \\
\hline
\end{tabular}
\label{tab:tab2}
\end{table}

%
%

\begin{table}
\caption{
Orbital parameters of the \pstaro system from \cite{Gozdziewski2003d} 
for the epoch of the first observation (JD=2,450,831.608). 
The mass of the central star is equal to $1.08~\mbox{M}_{\sun}$.
The orbital periods come from the FMA as reciprocals of the
proper mean motion frequencies.
}
\smallskip
\begin{tabular}{lrr}
\hline
Orbital parameter \hspace{5em}  & \ \ Planet b \ \ & \ \ Planet c \ \ \\
\hline
$\mbox{m}_2 \sin i$ [M$_{\idm{J}}$] \dotfill &  2.33 & 1.69 \\
a [AU] \dotfill & 0.82 & 2.78 \\
P [d] \dotfill &  263    &  1632  \\
e \dotfill     &  0.349 & 0.076 \\
$\omega$ [deg]\dotfill & 115.2 & 294.4  \\
$M$~[deg] \dotfill & 129.37 & 352.9 \\
\hline
\end{tabular}
\label{tab:tab3}
\end{table}

\end{document}